# STRONG GRAVITATIONAL LENSING BY SCHWARZSCHILD BLACK HOLES

G. S. Bisnovatyi-Kogan[1,2,3] and O. Yu. Tsupko[1,3]

*The properties of the relativistic rings which show up in images of a source when a black hole lies between the source and observer are examined. The impact parameters are calculated, along with the distances of closest approach of the rays which form a relativistic ring, their angular sizes, and their "magnification" factors, which are much less than unity.*

Keywords: *strong gravitational lensing: Schwarzschild black hole*

## 1. Introduction

One of the basic approximations in the theory of gravitational lensing [1, 2] is the approximation of weak lensing, i.e., small deflection angles. In the case of Schwarzschild lensing by a point mass, this approximation means that the impact parameters for incident photons are much greater than the Schwarzschild radius of the lensing system. In most astrophysical situations involving gravitational lensing, the weak lensing condition is well satisfied and it is possible to limit oneself to that case.

In some cases, however, strong lensing effects and those associated with large deflection angles are of interest. Some effects associated with taking the motion of photons near the gravitational radius of a black hole into account, have been discussed before [3, 4]. An explicit analytic expression for the deflection angle in a Schwarzschild metric have been derived in the strong field limit [5, 6]. In this paper we investigate effects arising from the strong distortion of isotropic radiation from a star by the gravitational field of a black hole.

(1) Space Research Institute, Russian Academy of Sciences, Russia; e-mail: gkogan@iki.rssi.ru
(2) Joint Institute for Nuclear Research, Dubna, Russia
(3) Moscow Engineering Physics Institute, Moscow, Russia; e-mail: tsupko@iki.rssi.ru

## 2. Multiple rings around a black hole

Let us consider the motion of a photon in the neighborhood of a black hole with a Schwarzschild metric. We shall work with a system of units in which the Schwarzschild radius $R_S = 2M$ ($G=1$, $c=1$), where $M$ is the mass of the black hole. The Schwarzschild metric is given by

$$ds^2 = g_{ik} dx^i dx^k = \left(1 - \frac{2M}{r}\right) dt^2 - \frac{dr^2}{1 - \frac{2M}{r}} - r^2 \left(d\theta^2 + \sin^2\theta \, d\varphi^2\right). \tag{1}$$

The equations for the radius $r$ and for the angular $\varphi$ and temporal $t$ coordinates, which determine the orbit of the photon in the Schwarzschild metric, can be written as [7]

$$\left(\frac{dr}{d\lambda}\right)^2 + B^{-2}(r) = b^{-2}, \tag{2}$$

$$\frac{d\varphi}{d\lambda} = \frac{1}{r^2}, \tag{3}$$

$$\frac{dt}{d\lambda} = b^{-1} \left(1 - \frac{2M}{r}\right)^{-1}. \tag{4}$$

Here $B^{-2}(r) = (1/r^2)(1 - (2M/r))$ is the effective potential and $b$ is the impact parameter. The shape of the orbit of a photon incident from infinity on a black hole is determined [7] by its impact parameter $b$. A detailed analysis of the photon orbits for all values of the impact parameter can be found elsewhere [8].

1. If $b < 3\sqrt{3} M$, then the photon falls to $R_s = 2M$ and is absorbed by the black hole.

2. If $b > 3\sqrt{3} M$, then the photon is deflected by an angle $\hat{\alpha}$ and flies off to infinity. Here there are two possibilities:

(a) If $b \gg 3\sqrt{3} M$, then the orbit is almost a straight line with a small deflection by an angle $\hat{\alpha} = 4M/R$, where $R$ is the distance of closest approach. This is the case customarily examined in the theory of weak gravitational lensing, when the impact parameter is much greater than the Schwarzschild radius of the lens.

(b) If $0 < b/M - 3\sqrt{3} \ll 1$, then the photon makes several turns around the black hole near a radius $r = 3M$ and flies off to infinity.

It is easy to determine the distance of closest approach $R$ to the black hole, defined by the condition $dr/d\lambda = 0$ and $b = B(R)$, using Eq. (2). The impact parameter b and the distance of closest approach R are related by

$$b^2 = \frac{R^3}{R - 2M}. \tag{5}$$

Obviously, the critical value $b = 3\sqrt{3} M$ corresponds to a distance of closest approach equal to $R = 3M$. In the case of large impact parameters, $b \gg M$, the deflection angle for a photon incident from infinity is $\hat{\alpha} = 4M/R$ [9] and, Eq. (5) implies that $b \approx R(1 + M/R)$, i.e., the impact parameter and the distance of closest approach are almost the same [8].

The exact deflection angle in a Schwarzschild metric can be derived from Eqs. (2) and (3). It is easy to see that

$$\frac{d\varphi}{dr} = \frac{1}{r^2 \sqrt{\frac{1}{b^2} - \frac{1}{r^2}\left(1 - \frac{2M}{r}\right)}}. \tag{6}$$

A rectilinear ray corresponds to a change in the angular coordinate of the trajectory by $\pi$, so that, Eq. (5) implies that the deflection angle of a photon as a function of the mass $M$ and radius $R$ of closest approach is

$$\hat{\alpha} = 2\int_R^\infty \frac{dr}{r^2 \sqrt{\frac{1}{b^2} - \frac{1}{r^2}\left(1 - \frac{2M}{r}\right)}} - \pi. \tag{7}$$

An algorithm for calculating the exact deflection angle $\hat{\alpha}(b/M)$ is given in Ref. 7. It is based on representing the integral in Eq. (7) as an elliptic integral

$$\int_R^\infty \frac{dr}{r^2 \sqrt{\frac{1}{b^2} - \frac{1}{r^2}\left(1 - \frac{2M}{r}\right)}} = 2\sqrt{\frac{R}{Q}} \, F\!\left(\sqrt{\frac{8Q}{(6+Q-R)(R-2+Q)}}, \sqrt{\frac{6+Q-R}{2Q}}\right), \tag{8}$$

where

$$Q^2 = (R-2)(R+6). \tag{9}$$

An expression for the exact deflection angle in terms of elliptical integrals was first given in Ref. 8. A detailed procedure for reducing the integral of Eq. (7) to an elliptic integral of the first kind, $F(z,k)$, is given in the Appendix. An abridged form of this procedure can be found in Ref. 8.

Thus, the exact deflection angle $\hat{\alpha}(b/M)$ is calculated as follows. [7] We represent an elliptic integral of the first kind in the form

$$F(z,k) = \int_0^z \frac{dt}{\sqrt{(1-t^2)(1-k^2 t^2)}} = \int_0^{\arcsin\phi_0} \frac{d\phi}{\sqrt{1-k^2 \sin^2\phi}}, \quad \sin\phi_0 = z. \tag{10}$$

1. Choose a value $r = R$ for the Schwarzschild coordinate of the point of closest approach.
2. Calculate the impact parameter using the formula $b^2 = R^3/(R-2)$.
3. Calculate $Q$ using the formula $Q^2 = (R-2)(R+6)$.
4. Determine the modulus of the elliptic integral of the first kind, $k$, using the formula $k^2 = (6+Q-R)/2Q$.
5. Determine from the relation $\sin^2\varphi_{min} = (2+Q-R)/(6+Q-R)$.
6. Then the total deflection will be equal to

$$\hat{\alpha} = 4(R/Q)^{1/2}[F(1,k) - F(\sin\varphi_{min}, k)] - \pi. \tag{11}$$

Note that if the value of the impact parameter is close to the critical value ($b \approx 3\sqrt{3}\,M$, $b > 3\sqrt{3}\,M$, $R - 3M \ll 1$), then a photon incident from infinity undergoes one or more revolutions around the black hole at $r \approx 3M$ and then escapes to infinity. In this case (the strong deflection limit) the deflection angle can be written in the form [5,8]

$$\hat{\alpha} = -2\ln\frac{R-3M}{36(2-\sqrt{3})M} - \pi. \tag{12}$$

The corresponding impact parameter $b$ can be found by expanding Eq. (5) in a series up to the quadratic terms in the small parameter $(R - 3M)$. The impact parameter is then equal to [8]

$$b = 3\sqrt{3}\,M + \frac{\sqrt{3}}{2}\frac{(R-3M)^2}{M}. \tag{13}$$

Using this expression, we can write the deflection angle as a function of the impact parameter $b$ in the form

$$\hat{\alpha} = -\ln(b/M - 3\sqrt{3}) + \ln[648(7\sqrt{3} - 12)] - \pi =$$
$$= -\ln\left(\frac{b}{b_{cr}} - 1\right) + \ln[216(7 - 4\sqrt{3})] - \pi \approx -\ln\left(\frac{b}{b_{cr}} - 1\right) - 0.40023, \tag{14}$$

where $b_{cr} = 3\sqrt{3}\,M$ is the critical value of the impact parameter that distinguishes photons which fall into the black hole from those which escape to infinity. This last expression for the deflection angle is the same as the one in Ref. 6.

Let the source, lens, and observer lie along a straight line (Fig. 1). The theory of gravitational lensing shows that in this case a circle, known as the Einstein ring, is formed [1]. Inside this "main" ring there are rings formed by photons which have been deflected by $2\pi, 4\pi, 6\pi, ...$; these rings are sometimes referred to as relativistic rings (Fig. 2). Equating $\hat{\alpha}$ to $2\pi, 4\pi, 6\pi, ...$, we find that these relativistic rings are localized at the impact parameters

$$b/M - 3\sqrt{3} = 0.00653, 0.0000121, 0.0000000227, 0.423 \cdot 10^{-10}, 0.791 \cdot 10^{-13}...$$

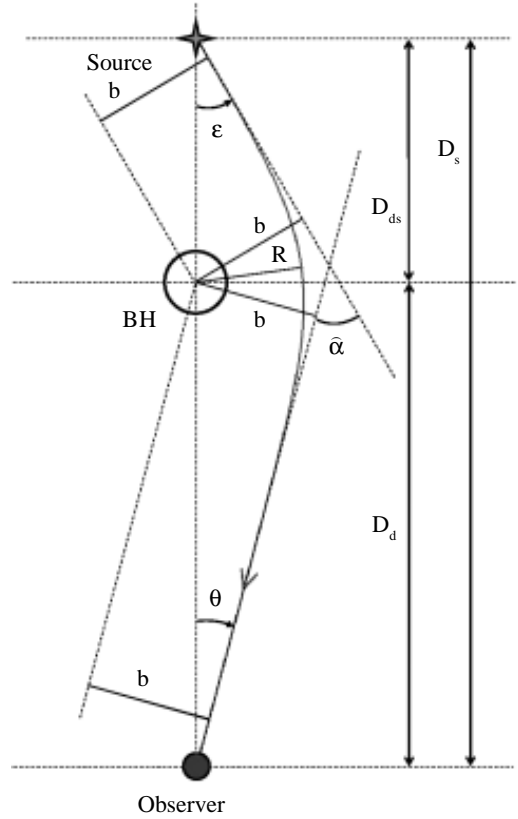

Fig. 1. The deflection of a ray of light when the source, lens, and observer lie on a straight line. The trajectory of the ray was calculated using Eqs. (2) and (3) for the following values of the parameters: $R_s = 2$, $R = 7$, $b \approx 8.3$, $D_d = 32$, and $D_{ds} = 16$.

The first two terms are given in the book by Misner et al. [7] The corresponding distances of closest approach are

$$R/M - 3 = 0.0902, 0.00375, 0.000162, 0.699 \cdot 10^{-5}, 0.302 \cdot 10^{-6}, ...$$

It must be noted that, in reality, for the case in which the source, observer, and lens lie on a single straight line, a photon must be deflected by an angle that is not exactly equal to $2\pi n$, but slightly greater, so that $b$ and $R$ are smaller by a small correction which we shall ignore, as is done in Ref. 7. The fact that the deflection angle differs from $2\pi n$ is taken into account in deriving the formulas for the flux from the relativistic rings, where these corrections are of fundamental importance, since they determine the thickness of a ring owing to the fact that the rays are deflected at different angles.

The values of the impact parameters can also be obtained using the strong field approximation, i.e., Eq. (14). Using the equation $\hat{\alpha}(b) = 2\pi n$ ($n = 1, 2, 3, ...$), we obtain the following values of $b_n$:

$$b_n = b_{cr}\left(1 + e^{C_1 - 2\pi n}\right), \quad C_1 = -0.40023. \tag{15}$$

Calculations with this formula show that the strong field approximation yields an accuracy of about 0.4% for the first relativistic ring and better than that for the other rings.

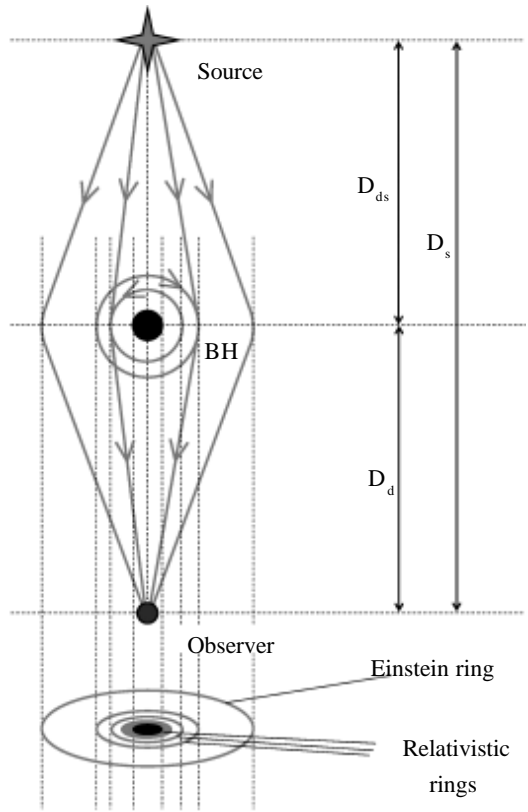

Fig. 2. The Einstein ring and relativistic rings.

The angular size of a ring (the angle between the lines passing through the observer and the lens (black hole) and the direction of the image of the source (ring)) is given by $\theta = b/D_d$, where $D_d$ is the distance between the observer and the lens [1]. We now find the angular size $\theta_0$ of the main ring. Figure 1 shows that the following relationship exists between the angles:

$$\theta_0 + \varepsilon = \hat{\alpha}. \tag{16}$$

The impact parameter b is related to the angles $\theta_0$ and $\varepsilon$ by

$$b = D_{ds}\, \varepsilon = D_d\, \theta_0\,; \quad \varepsilon = \frac{D_d}{D_{ds}} \theta_0. \tag{17}$$

Now we express the deflection angle $\hat{\alpha}$ in terms of the angle $\theta_0$, using the weak deflection approximation, $b \simeq R \gg 2M$:

$$\hat{\alpha} = \frac{4M}{R} \simeq \frac{4M}{b} = \frac{4M}{D_d} \frac{1}{\theta_0}. \tag{18}$$

Substituting for the angle $\varepsilon$ from Eq. (17) and for the deflection angle (18) in Eq. (16) we obtain an equation for $\theta_0$:

$$\theta_0 + \frac{D_d}{D_{ds}} \theta_0 = \frac{4M}{D_d} \frac{1}{\theta_0}. \tag{19}$$

Then, given that $D_d + D_{ds} = D_s$, we obtain

$$\theta_0 = \sqrt{4M \frac{D_{ds}}{D_d D_s}}, \tag{20}$$

where $D_s$ is the distance from the observer to the source and $D_{ds}$ is the distance between the lens and the source. The angular sizes $\theta_n$ of the relativistic rings are given by Eqs. (17) and (15), together with the relationships $\hat{\alpha} = 2\pi n$ and $b = b_{cr}\left(1 + e^{C_1 - 2\pi n}\right)$ for the relativistic rings, as

$$\theta_n = \frac{b_{cr}}{D_d}\left(1 + e^{C_1 - 2\pi n}\right). \tag{21}$$

In the simplest model of a point Schwarzschild gravitational lens (in the weak deflection approximation) the source is treated as a point. If the source, lens, and observer do not lie on a straight line, then two images of the source are formed, and their angular positions and flux magnifications relative to an unlensed source can be calculated [1]. Besides the two images obtained in the weak field limit, two sequences of images of a point source are formed on different sides of the lens owing to photons which undergo one or more turns around the lens. The positions of these images can be calculated using the exact expression for the strong field limit, as well as the exact expression for the deflection angle. The strong field limit makes it possible to calculate both the angular position of the images and their magnifications [5,8]. On the other hand, in the case where the source, lens, and observer lie on a straight line, treating the source as a point is not suitable, since that leads to a divergence (infinite magnification). To calculate the flux gain in this case it is necessary to consider a source of finite size, rather than a point source. Given that the source is circular, with a uniform surface luminosity and a specified finite angular size, the flux gain can be calculated [1]. The fluxes from the relativistic rings resulting from strong lensing of the source can be treated in a similar way.

The flux from the source is determined by its surface brightness and the solid angle it occupies in the sky. Since the gravitational deflection of light does not involve emission or absorption, the intensity of the radiation remains constant along a beam. In addition, the gravitational deflection of light by a local, essentially static lens does not introduce an additional shift in frequency. Thus, the surface brightness of an image of a lensed source is equal to the brightness of the image in the absence of the lens. Consequently, the ratio of the fluxes in the images of a lensed and a nonlensed source is determined by the ratio of the solid angles of these sources in the sky [1], i.e., by the ratio of the areas of the corresponding images on a photographic plate or CCD array (in the case of ideal imaging). The ratio

$$\mu = \frac{\Delta\omega}{(\Delta\omega)_0}$$

where $(\Delta\omega)_0$ and $(\Delta\omega)$, respectively, denote the visual solid angles of the unlensed and lensed sources, is referred to as the magnification factor.

We now find the magnification factor for the main image of a circular source with angular radius $\beta$ and uniform brightness lying in a straight line with a lens and an observer. The image of the source is an annulus. We introduce $y = \beta/\theta_0$, the ratio of the angular size of the source to the angular size of the main ring. The unlensed source occupies a solid angle of $(\Delta\omega)_0 = \pi\beta^2 = \pi y^2 \theta_0^2$. It can be shown [1] that in the weak lensing limit, a circular source transforms into a ring with the following inner and outer angular radii: $\theta_{in} = \frac{\theta_0}{2}\left(\sqrt{y^2+4} - y\right)$ and $\theta_{out} = \frac{\theta_0}{2}\left(\sqrt{y^2+4} + y\right)$. Thus, the solid angle occupied by the ring is

$$\Delta\omega = \pi(\theta_{out}^2 - \theta_{in}^2) = \pi\frac{\theta_0^2}{4}\left[\left(\sqrt{y^2+4}+y\right)^2 - \left(\sqrt{y^2+4}-y\right)^2\right] = \pi\theta_0^2 y\sqrt{y^2+4}. \tag{22}$$

Thus, the magnification factor $\mu_0$ for the main ring is

$$\mu_0 = \frac{\sqrt{y^2+4}}{y} \simeq \frac{2}{y}, \quad \text{for} \quad y \ll 1. \tag{23}$$

We now find the flux "magnification" for the relativistic rings of this source. It can be shown, with using of Ref. 10, that the $n$-th relativistic ring will have the following inner and outer angular radii:

$$\theta_{in}^n = \theta_n(1 - A_n(\theta_n + \beta)), \quad \theta_{out}^n = \theta_n(1 - A_n(\theta_n - \beta)), \tag{24}$$

where

$$A_n = \frac{D_s}{D_{ds} D_d} \frac{b_{cr}}{\theta_n} e^{C_1 - 2\pi n}.$$

The solid angle occupied by this relativistic ring is

$$\Delta\omega = \pi\left[(\theta_{out}^n)^2 - (\theta_{in}^n)^2\right] = 4\pi\theta_n^2(1 - A_n\theta_n)A_n\beta. \tag{25}$$

Thus, the "magnification" factors $\mu_n$ of the relativistic rings are given by

$$\mu_n = \frac{4\theta_n^2(1 - A_n\theta_n)A_n}{\beta}. \tag{26}$$

The term

$$A_n \theta_n = \frac{D_s}{D_{ds} D_d} b_{cr} e^{C_1 - 2\pi n} \ll 1, \qquad (27)$$

so that, to great accuracy, we have

$$\mu_n = 4 \frac{\theta_n}{\beta} A_n \theta_n. \qquad (28)$$

After some transformations, we obtain

$$\mu_n = 4 \frac{b_{cr}^2}{\beta} \frac{D_s}{D_{ds} D_d^2} \left(1 + e^{C_1 - 2\pi n}\right) e^{C_1 - 2\pi n} \ll \mu_0. \qquad (29)$$

Using the expressions for $b_{cr} = 3\sqrt{3} M = (3\sqrt{3}/2) R_S$ and $\beta = R_*/D_s$, where $R_*$ is the source radius, we obtain

$$\mu_n = 27 \frac{R_S^2 D_s^2}{R_* D_{ds} D_d^2} \left(1 + e^{C_1 - 2\pi n}\right) e^{C_1 - 2\pi n} \ll 1. \qquad (30)$$

For a distant quasar with $M_* = 10^9 M_\odot$, $R_* = 15 R_{*S} (R_{*S} = 2GM_*/c^2)$, $D_{ds} = 10^3$ Mpc, $D_d = 3$ Mpc, $D_s \approx D_{ds}$, and a lens of mass $M = 10^7 M_\odot$, we obtain

$$\mu_0 = 1 \cdot 10^6, \quad \mu_1 = 2 \cdot 10^{-15}, \quad \mu_2 = 4 \cdot 10^{-18}. \qquad (31)$$

For the same quasar with a lens of mass $M = 20 M_\odot$ and $D_d = 1$ kpc, we obtain

$$\mu_0 = 9 \cdot 10^4, \quad \mu_1 = 9 \cdot 10^{-20}, \quad \mu_2 = 2 \cdot 10^{-22}. \qquad (32)$$

The large magnification $\mu_0$ follows from the small dimensions of the source and the assumption that the source, lens, and observer are on a straight line. For a point source lying on a straight line with the observer and the lens, the magnification goes to infinity because of the assumption that the source is a point and the use of the approximation of geometric optics. If, on the other hand, we consider a source with a finite angular size, rather than point sources, then the infinite magnification becomes finite, but increases as the source size decreases. On the other hand, for small sources, it becomes important to account for various effects, primarily wave effects, which actually lower the magnification factor [2].

A general formula for the magnification factor of a finite diameter sources is given in Ref. 5. The approximation of deflection by an angle $2\pi n$ is used there, while small deviations from these angles are neglected, and the magnification coefficient is found by integrating the formulas for point sources, rather than by the "geometric" approach employed here. Equation (29) is the same as the expression which can be obtained from the general formula for a source of finite size in Ref. 5.

## 3. Radiation diagram for a point source located near a black hole

Let us examine in more detail the influence of the field of a black hole lying on a straight line with the observer, on the angular distribution of the radiation of a star (Fig. 1). Let us draw an arbitrary plane passing through the star perpendicular to the line $D_{ds}$. We shall say that emitted photons have flown "upward" if they end up in the upper half plane at a large distance from the star. The photons have flown "downward" if the end up in the lower half plane. In the absence of a black hole, the star radiates isotropically. Thus, equal fractions of the radiation go "up" and "down" (Fig. 3a, where the photons moving "upward" and "downward" are indicated by different shading).

When a black hole lies "under" the star between the star and the observer, the angular distribution of the radiation changes. We shall take three effects into account:

1) capture by the black hole of photons emitted with small impact parameters;

2) deflection of photons emitted at angles near 90° from vertical; and,

3) the "upward" return of photons that were emitted downward and have undergone a half turn around the black hole.

A calculation has been done for the following system parameters: distances (in mass units) $D_{ds} = 10000\,M$ and Schwarzschild radius $R_S = 2M$. For photons emitted almost perpendicular to the axis, we used the approximation of weak deflection, $\hat{\alpha} = 2M/b = 2M/D_{ds}$.[1] For the photons that are emitted "downward," we used another approximation: suppose that they are emitted at infinity from a black hole (and not at a distance $D_{ds}$), which is the standard approximation for gravitational lensing. Because of this, we assume that the impact parameter $b$ (the impact parameter at infinity) is equal to the "geometric" impact parameter for emission of photons by the star, $b = D_{ds} \sin\varepsilon \simeq D_{ds}\,\varepsilon$, where $\varepsilon$ is the emission angle reckoned from the vertical (Fig. 1). For the returning photons, we used the algorithm for calculating the deflection taken from Ref. 7 and described above.

The results of the calculation are shown in Fig. 3b. The black region represents the range of emission angles from which photons are captures by the black hole. It is also evident that some of the photons initially emitted "upward," turn "downward" and that some of the photons emitted "downward," return "upward." Three characteristic angular regions are indicated in the figure: A, B, and C. Region A (because of symmetry, there are two such regions in the figure) is the angular range for photons emitted at angles greater than 90° from the vertical, but which move "down" because of they are attracted by the black hole. Region B (there are two such regions in the figure) is the angular range for photons emitted at angles close to the critical value (for capture) which undergo a half turn and escape "upwards." Region C (indicated in black in the figure) is the angular range for the emitted photons that are captured by the black hole. The angular size of region A=0°.0115, of region B=0°.0054, and region C=0°.0298. In the figure the sizes of all three regions are greatly enhanced for visibility, but the ratios of the angular sizes of regions A, B, and C are maintained.

The calculations yield two qualitative results, which hold true for other distances $D_{ds}$ between the black hole and the star (when the same approximations are used for the deflection angles):

1. When a black hole is present, more will be radiated "upward" because of the large fraction of photons captured by the black hole.

2. The number of photons emitted nearly perpendicular to the vertical and deflected "downward" is greater than the number of photons which undergo a half turn and escape "upward."

Of course, only a small fraction of the rays deflected into a given half plane reach the observer. If the observer is in a straight line with a black hole and a star in the "lower" half plane, then the deflection of photons emitted at angles slightly greater than 90 degrees from the vertical will generally have no effect on the observations.

---

1 The deflection angle is $\hat{\alpha} = 4M/b$ if a photon is incident from infinity, reaches the distance of closest approach, and flies off to infinity. In this case, the photon traverses half this path-- it moves from the point of closest approach to infinity. Thus, for such a photon, $\hat{\alpha} = 2M/b$.

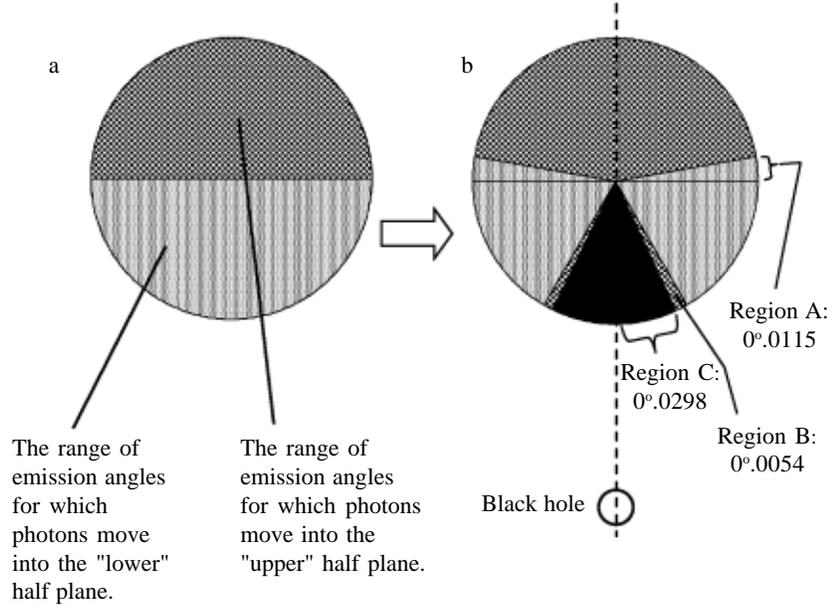

Fig. 3. A black hole deflects light emitted by an isotropically radiating source. The emission direction of a photon is specified by the angle $\varepsilon$ relative to the vertical; $\varepsilon = (0, \pi)$.

**Appendix**

We examine the two integrals $\alpha^{(1)}$ and $\alpha^{(2)}$, and show that they are identically equal to one another.

$$\alpha^{(1)} = 2\int_R^\infty \frac{dr}{r^2 \sqrt{\frac{1}{b^2} - \frac{1}{r^2}\left(1 - \frac{2}{r}\right)}}, \quad \text{where} \quad b^2 = R^3/(R-2); \tag{33}$$

and

$$\alpha^{(2)} = 4(R/Q)^{1/2}[F(1, k) - F(\sin\varphi_{min}, k)], \tag{34}$$

where

$$Q^2 = (R-2)(R+6), \quad k^2 = (Q-R+6)/2Q, \quad \sin^2\varphi_{min} = (2+Q-R)/(6+Q-R).$$

In $\alpha^{(1)}$ make the substitutions $u = 1/r$ and $du = (-1/r^2)dr$. With this transformation, $R \to 1/R, \infty \to 0$, so that

$$2\int_R^\infty \frac{dr}{r^2\sqrt{\frac{1}{b^2}-\frac{1}{r^2}\left(1-\frac{2}{r}\right)}} = -2\int_{1/R}^0 \frac{du}{\sqrt{\frac{1}{b^2}-u^2(1-2u)}} = 2\int_0^{1/R} \frac{du}{\sqrt{2u^3-u^2+\frac{1}{b^2}}} =$$

$$= 2\int_0^{1/R} \frac{du}{\sqrt{2u^3-u^2+\frac{R-2}{R^3}}} = 2R^{3/2}\int_0^{1/R} \frac{du}{\sqrt{2u^3R^3-u^2R^3+R-2}}.$$

The expression under the square root sign is a cubic polynomial in $u$ and we shall decompose it into three cofactors. To do this we have to find the roots of the cubic equation $2u^3R^3-u^2R^3+R-2=0$. One of them is easy to find by taking $u=1/R$. Knowing this root, we can factor the cubic polynomial into the product

$$2u^3R^3-u^2R^3+R-2 = (u-1/R)\left[2R^3u^2-(R-2)R^2u-(R-2)R\right].$$

The quadratic polynomial in square brackets can be factored by separating the complete square:

$$2R^3u^2-(R-2)R^2u-(R-2)R = 2R^3(u-A)(u-C).$$

Here

$$A = \frac{R-2+Q}{4R}, \quad C = \frac{R-2-Q}{4R} < 0,$$

where

$$Q^2 = R^2+4R-12 = (R-2)(R+6).$$

Thus, we have reduced the expression under the root sign to the product of three cofactors, i.e.,

$$2u^3R^3-u^2R^3+R-2 = 2R^3(u-A)(u-B)(u-C), \quad \text{where} \quad B \equiv 1/R.$$

The following formula is given in Ref. 11 (the upper limit is equal to the constant in one of the factors):

$$\int_u^b \frac{dx}{\sqrt{(x-a)(x-b)(x-c)}} = \frac{2}{\sqrt{a-c}} F(\delta,q), \quad [a > b > u \geq c],$$

where

$$\delta = \arcsin\sqrt{\frac{(a-c)(b-u)}{(b-c)(a-u)}}, \quad q = \sqrt{\frac{b-c}{a-c}}.$$

In our case

$$\frac{(A-C)B}{(B-C)A} = \frac{8Q}{(6-R+Q)(R-2+Q)}, \quad \frac{B-C}{A-C} = \frac{6-R+Q}{2Q}, \quad A-C = \frac{Q}{2R}.$$

Hence,

$$\int_0^{1/R} \frac{du}{\sqrt{(u-A)(u-B)(u-C)}} = 2\sqrt{\frac{2R}{Q}} F\left(\sqrt{\frac{8Q}{(6-R+Q)(R-2+Q)}}, \sqrt{\frac{6-R+Q}{2Q}}\right).$$

Therefore, we have reduced the original integral to an elliptic integral of the first kind. The equation given in

Eq. (7) includes the difference of two elliptic integrals (see Eq. (34)), one of which is the so-called complete elliptic integral (upper limit =1). It turns out that the integral obtained above is the same as this difference.

The following property of elliptic integrals is given in Ref. 12 (we cite the elliptic integral (10) in our notation):

$$F(z_1, k) + F(z_2, k) = F(1, k),$$

if the following relationship of the arguments is satisfied:

$$\sqrt{1-k^2} \frac{z_1}{\sqrt{1-z_1^2}} \frac{z_2}{\sqrt{1-z_2^2}} = 1.$$

Appropriate calculations show that for $z_1 = \sqrt{\frac{8Q}{(6-R+Q)(R-2+Q)}}$, $z_2 = \sqrt{\frac{2+Q-R}{6+Q-R}}$, and $k = \sqrt{\frac{6-R+Q}{2Q}}$ this equation is satisfied. Hence,

$$F\left(\sqrt{\frac{8Q}{(6-R+Q)(R-2+Q)}}, \sqrt{\frac{6-R+Q}{2Q}}\right) =$$
$$= F\left(1, \sqrt{\frac{6-R+Q}{2Q}}\right) - F\left(\sqrt{\frac{2+Q-R}{6+Q-R}}, \sqrt{\frac{6-R+Q}{2Q}}\right).$$

And so, we write the complete transformation in the form

$$\alpha^{(1)} = 2 \int_R^\infty \frac{dr}{r^2 \sqrt{\frac{1}{b^2} - \frac{1}{r^2}\left(1 - \frac{2}{r}\right)}} = 2 \int_0^{1/R} \frac{du}{\sqrt{2u^3 - u^2 + \frac{1}{b^2}}} = 2R^{3/2} \int_0^{1/R} \frac{du}{\sqrt{2u^3 R^3 - u^2 R^3 + R - 2}} =$$

$$= 2R^{3/2} \int_0^{1/R} \frac{du}{\sqrt{(u - 1/R)(2R^3 u^2 - (R-2)R^2 u - (R-2)R)}} =$$

$$= \frac{2R^{3/2}}{\sqrt{2} R^{3/2}} \int_0^B \frac{du}{\sqrt{(u-A)(u-B)(u-C)}} =$$

$$= \sqrt{2} \, 2 \sqrt{\frac{2R}{Q}} F\left(\sqrt{\frac{8Q}{(6-R+Q)(R-2+Q)}}, \sqrt{\frac{6-R+Q}{2Q}}\right) =$$

$$= 4\sqrt{\frac{R}{Q}} F\left(\sqrt{\frac{8Q}{(6-R+Q)(R-2+Q)}}, \sqrt{\frac{6-R+Q}{2Q}}\right) =$$

$$= 4\sqrt{\frac{R}{Q}} \left[F\left(1, \sqrt{\frac{6-R+Q}{2Q}}\right) - F\left(\sqrt{\frac{2+Q-R}{6-R+Q}}, \sqrt{\frac{6-R+Q}{2Q}}\right)\right] =$$

$$= 4(R/Q)^{1/2} [F(1, k) - F(\sin\varphi_{min}, k)] = \alpha^{(2)}.$$

This work was partially supported by RFBR grants 05-02-17697, 06-02-90864 and 06-02-91157, RAN Program "Formation and evolution of stars and galaxies" and Grant for Leading Scientific Schools NSh-10181.2006.2.

# REFERENCES

1. P. Schneider, J. Ehlers, and E. Falco, Gravitational Lensing, Springer-Verlag, Berlin (1992).
2. P. V. Bliokh and A. A. Minakov, Gravitational Lenses [in Russian], Naukova Dumka, Kiev (1989).
3. W. Ames and K. Thorne, *Astrophys. J.* **151**, 659 (1968).
4. G. S. Bisnovatyi-Kogan and A. A. Ruzmaikin, *Astrophys. Space Sci.* **28**, 45 (1974).
5. V. Bozza, S. Capozziello, G. Iovane, and G. Scarpetta, *General Relativity and Gravitation* **33**, 1535 (2001).
6. V. Bozza, *Phys. Rev. D* **66**, 103001 (2002).
7. C. W. Misner, K. S. Thorne, and J. A. Wheeler, Gravitation, Freeman, New York (1973).
8. C. Darwin, Proceedings of the Royal Society of London, Series A, *Mathematical and Physical Sciences* **249** (1257), 180 (1959).
9. L. D. Landau and E. M. Lifshitz, Field Theory [in Russian], Nauka, Moscow (1962).
10. V. Bozza and M. Sereno, *Phys. Rev. D* **73**, 103004 (2006).
11. I. S. Gradshtein and I. M. Ryzhik, Tables of Integrals, Sums, and Products [in Russian], Nauka, Moscow (1971).
12. M. Abramowitz and I. Stegun, Handbook of Special Functions with Formulas, Graphs, and Mathematical Tables [Russian translation], Nauka, Moscow (1979).